\documentstyle[preprint,aps]{revtex}
\def\sgommd{$\sigma$-$\omega$ model~}
\def\vr{\bf r}

\newcommand{\psla}{\mbox{\ooalign{\hfil/\hfil\crcr$p$}}}

\newcommand{\delsla}{\mbox{\ooalign{\hfil/\hfil\crcr{$\partial$}}}}

\newcommand{\fpisq}{\mbox{$f_{\pi}^2$}}
\newcommand{\mpi}{\mbox{$m_{\pi}$}}
\newcommand{\mpisq}{\mbox{$m_{\pi}^2$}}
\newcommand{\eq}{\label}

\newcommand{\vp}{\mbox{\boldmath $p$}}
\newcommand{\vk}{\mbox{\boldmath $k$}}

\newcommand{\psibar}{\mbox{${\bar \psi}$}}

\newcommand{\Pitild}{\mbox{${\tilde{\Pi}}$}}

\newcommand{\vPi}{\mbox{${\boldmath \Pi}$}}
\newcommand{\rhtil}{\mbox{${\tilde{\rho}}$}}

\newcommand{\epsi}{\mbox{$\varepsilon$}}

\newcommand{\stret}[1]{\hbox{$\vcenter to #1{}$}}

\def\sigm{\mbox{$\langle \sigma \rangle$}}
\def\gs{\mbox{$g_\sigma$}}
\def\gv{\mbox{$g_\omega$}}
\def\ms{\mbox{$m_\sigma$}}
\def\mv{\mbox{$m_\omega$}}

\begin{document}

\title{\bf  
Isoscalar Giant Quadrupole Resonance State
in the Relativistic Approach 
with the Momentum-Dependent Self-Energies}

\author{Tomoyuki~Maruyama$^{1,2}$ 
and Satoshi~Chiba$^{3}$
\\[1ex]  
$^{1}$ College of Bioresource Sciences \\
Nihon University \\
Fujisawa, Kanagawa 252-8510, Japan \\[1ex]
$^{2}$ The Institute of Physical and Chemical Research (RIKEN)\\
Hirosawa 2-1, Wako-shi, Saitama 351-0198, Japan \\[1ex]
$^{3}$ Advanced Science Research Center, \\
Japan Atomic Energy Research Institute,\\
Tokai, Ibaraki 319-1195, Japan }

\maketitle

\begin{abstract}
We study the excited energy of the isoscalar giant quadrupole
resonance with the scaling method in the relativistic
many-body framework.
In this calculation we introduce the momentum-dependent parts
of the Dirac self-energies arising from the one-pion exchange on the
assumption of the pseudo-vector coupling with nucleon field.
It is shown that this momentum-dependence enhances the
Landau mass significantly and thus suppresses
the quadrupole resonance energy even giving 
the small Dirac effective mass which causes a problem in
the momentum-independent mean-field theory.
\end{abstract}


\vfil
\eject

\newpage

The past decades have seen many successes in the relativistic 
treatment of the nuclear many-body problem.
The relativistic framework has big advantages in several 
aspects \cite{Serot}: a useful Dirac phenomenology for the 
description of nucleon-nucleus scattering \cite{Hama,Tjon}, 
the natural incorporation of the spin-orbit force \cite{Serot} 
and the structure of extreme nuclei \cite{hirata}.
These results have shown that there are large attractive scalar 
and repulsive vector fields, and that the nucleon 
effective mass becomes small in the medium.

The small effective mass enlarges the Fermi velocity, and it
causes troubles in some observables.  For example, the 
isoscalar giant quadrupole resonance (ISGQR) state 
is predicted at a too high excitation energy due to the 
large Fermi velocity\cite{Nishi}.
In this subject, it is assumed that  
the momentum-dependence of the Dirac fields is negligible 
in the low energy region, particularly below the Fermi level.
In fact, only very small momentum-dependence has appeared in 
the relativistic Hartree-Fock (RHF) calculation 
though the Fock contribution is not small \cite{RHF,soutome}.
Thus the Fock contributions are thought to be incorporated
into the relativistic Hartree (RH) approximation by 
introducing complicated density-dependent
interaction and fitting parameters \cite{hirata,RHF,Cheng}.

In the high energy region the vector-fields must become very small 
to explain the optical potential of the proton-nucleus elastic 
scattering 
\cite{Hama,KLW1} and the transverse flow in the heavy-ion collisions 
\cite{TOMO1}.
A Dirac-Bruckner-Hartree-Fock (DBHF) calculation 
has also shown that the 
momentum-dependence changes the nuclear equation of state noticeably
\cite{DBHF2}.
Furthermore, Weber et al. \cite{KLW1} have suggested 
that the Fermi velocity does not correspond to
the effective mass uniquely when introducing the momentum-dependence
of the Dirac fields.

We can easily suppose that it is the one-pion exchange force 
which  
produces the major momentum-dependence because the 
interaction range 
is largest.
In this paper, thus, we introduce the momentum-dependence
to the Dirac fields due to the one-pion exchange, and 
discuss how the Fock parts 
given by the one-pion exchange affects the excitation
energy of ISGQR.
Actually we use the scaling method in the way of 
Ref.\cite{Nishi} which is proved to give consistent
results with RPA for the giant
multipole states in the non-relativistic framework\cite{bohi}.
This relation has been confirmed also in the relativistic
framework for the monopole vibration mode\cite{tomoGR}.

Let us consider the infinite nuclear matter system 
with the isospin symmetry.
The nucleon propagator in the self-energy $\Sigma$ is 
given by 
\begin{equation} 
S^{-1}(p) = \psla - M - \Sigma(p),
\label{prop}
\end{equation} 
where $\Sigma(p)$ has a Lorentz scalar part $U_{s}(p)$ and 
a Lorentz vector part $U_{\mu}(p)$ as
\begin{equation}
\Sigma(p) = - U_{s}(p) + \gamma^{\mu} U_{\mu}(p).
\end{equation} 
For the future convenience we define the effective mass and 
the kinetic momentum as follows.
\begin{eqnarray}
M^{*}(p) & = & M - U_{s}(p) ,
\nonumber \\
\Pi_{\mu}(p) & = & p_{\mu} - U_{\mu}(p) .
\end{eqnarray}
Using the on-mass-shell condition that 
${\Pi}^2(p) - M^{*2}(p) = 0$,
the single particle energy with momentum {\vp}  is defined as
\begin{equation}
\epsi({\vp})  =  p_0 |_{on-mass-shell}
= \sqrt{ \Pi^2({\vp}) + M^{*2}({\vp}) } + U_{0}(\vp) . 
\end{equation}
 
Next we consider the variation of the total energy
in the quadrupole deformation in order to discuss
ISGQR with the scaling method.
Using the scaling method, first, we vary the density-distribution
from the normal nuclear matter distribution $\rhtil_0 ({\vr})$
as
\begin{equation}
\rhtil_0 ({\vr}) \rightarrow \rhtil_\lambda ({\vr})
= \rhtil_0 ( e^{- \lambda} x, e^{- \lambda}y, e^{2\lambda}z ).
\end{equation}
In the uniform nuclear matter it is equivalent to
the variation of the momentum-distribution $n(\vp)$ as
\begin{equation}
n_0 (\vp) \rightarrow n(\vp) =  n_{\lambda}({\vp}) 
= n_{0}({\vp}_\lambda) = 
n_{0}( e^{\lambda}p_x, e^{\lambda}p_y, e^{-2\lambda}p_z ),
\end{equation}
where $n_0(\vp) = \theta(p_F - |\vp|)$  with the Fermi-momentum
$p_F$ at the saturation density $\rho_0$.

With the variation of the momentum-distribution as
$n_0(\vp) \rightarrow n_0(\vp) + {\delta}n_{\vp}$,
the single-particle energy $\epsi({\vp})$
is obtained in the Hartree-Fock framework by
\begin{equation}
\epsi({\vp}) = \frac{\delta E_T}{\delta n_{\vp}} = 
\sqrt{ \vPi(p)^2 + M^{* 2}(p) } + U_0(p)  
|_{p_0 = \epsi({\vp})},
\end{equation} 
>From this relation we get the following equations
\begin{eqnarray}
\frac{\partial E_T}{\partial \lambda} |_{\lambda = 0}
& = &   4 \Omega \int \frac{{\rm d}^3 {\vp}}{(2 \pi)^3} 
\{ \frac{\partial n({\vp})}{\partial \lambda} |_{\lambda = 0} \}
\epsi({\vp}) ~~=~~ 0
\\
\frac{\partial^2 E_T}{\partial \lambda^2} 
|_{\lambda = 0}
& = &   4 \Omega \int \frac{{\rm d}^3 {\vp}}{(2 \pi)^3} 
\{ \frac{\partial n({\vp})}{\partial \lambda} |_{\lambda = 0}\}
\{ D_\lambda \epsi({\vp})|_{\lambda = 0} \}
\label{sec-der}
\end{eqnarray}
The derivative of the single particle energy $\epsi$
can be written as
\begin{equation}
D_\lambda \epsi({\vp}_\lambda,\lambda)
= \frac{\partial {\vp}_\lambda}{\partial \lambda}
{\nabla}_{{\bf p}_\lambda} \epsi + 
\frac{\partial \epsi}{\partial \lambda}
\label{ederlam}
\end{equation}
where the total derivative ${\nabla}_{\bf p}$ is defined 
on the on-mass-shell condition: $p_0 = \epsi ({\vp})$.
In this equation the second term of the right-hand side 
$\partial \epsi /\partial \lambda$ 
corresponds to the derivative with the variation of 
the self-energies at the fixed momentum by changing 
the momentum-distribution.
This term holds the spherical symmetry at the limit of 
$\lambda \rightarrow 0$, and does not contribute to
the integral of the left-hand side in eq.(\ref{sec-der}).

Substituting eq.(\ref{ederlam}) into eq.(\ref{sec-der}), 
hence,
the restoring force of ISGQR $C_{Q}$ becomes
\begin{eqnarray}
C_{Q} & = &
\frac{\partial^2 {E_T}/A}{\partial \lambda^2} 
|_{\lambda = 0}
~=~  -  \frac{4 \Omega}{A} \int \frac{{\rm d}^3 {\vp}}{(2 \pi)^3} 
\delta(|\vp| - p_F) (\frac{\vp^2 - 3p_z^2}{|\vp|}) 
(\frac{\vp^2 - 3p_z^2}{M^{*}_L}) ,
\nonumber \\
& = & \frac{12}{5} \frac{{p_F}^2}{ M^{*}_L},
\end{eqnarray}
where
\begin{eqnarray}
M^{*}_{L} & = &
(2 \frac{d}{d \vp^2} \epsi({\vp}) )^{-1}
|_{|{\vp}| = p_{F}} ,
\end{eqnarray}
which is so called the '{\bf Landau mass}' corresponding
to  the effective mass in the non-relativistic framework.

In Ref.\cite{Nishi} the mass-parameter of ISGQR is given as
\begin{eqnarray}
B_{Q} & = & 2 {\epsi_F} < {\bf r}^2 > ,
\end{eqnarray} 
with the Fermi energy $\epsi_F$, and then
the frequency of ISGQR is obtained as
\begin{equation}
\omega_{Q} = \sqrt{ \frac{C_{Q}}{B_{Q}} }
 =  \sqrt{ \frac{6 {p_F}^2 }{ 5 M^*_L \epsi_F < {\bf r}^2 > }} 
\label{exenQ}
\end{equation}
This expression is same as that of Ref.\cite{Nishi} with RH 
except the Landau mass; 
$M^{*}_{L} = \sqrt{ {p_F}^2 + M^{* 2} }$ in RH.

At the saturation density,
the Fermi energy agrees with the total energy per
nucleon whose value is almost the same as that of nucleon mass: 
$\epsi_F = E_{T}/A \approx M$, 
In addition, nuclear radii are scaled to be proportional to $A^{1/3}$
as $ < r^2 >  =  3/5 r_0^2 A^{2/3} $
and then we get the frequency $\omega_Q$ as
\begin{equation}
\omega_Q \approx \sqrt{ \frac{4 <T^{nr}_K> }{ M^*_L  r_0^2  } } A^{-1/3} , 
\label{exenQ2}
\end{equation}
where $<T^{nr}_K>$ is the non-relativistic averaged kinetic energy as
\begin{equation}
<T^{nr}_K> ~=~ < \frac{\vp^2}{2 M} > ~=~ \frac{3 p_F^2}{10M}. 
\end{equation}
This expression completely coincides with that of the non-relativistic
model \cite{KA}

We substitute the empirical experimental values
$\omega_{Q} \approx 63 A^{-1/3} {\rm MeV}$, $<T_K^{nr}> \approx 25 {\rm MeV}$
and $r_0 \approx 1.125 {\rm fm}$ into the
eq.(\ref{exenQ2}) and obtain the value of 
the Landau mass  \cite{ISGQR} as
\begin{eqnarray}
M^{*}_{L} / M & \approx & 0.85 .
\end{eqnarray}

Consequently we cannot find any difference in the expression of ISGQR
between the relativistic and non-relativistic frameworks.
The main problem is whether we can give the above value of $M_L^*$
with holding consistency to other observables;
the usual analyses indicate that $M^*/M = 0.55 - 0.7$, which
gives $M_L^{*} \approx 0.6 - 0.75$ in RH. 


As a next step we explain the details of our calculation in this work.
We consider a model of the usual \sgommd plus 
the Fock part of the one-pion exchange. 

Along this line
we define a Lagrangian density in the system as
\begin{eqnarray}
{\cal L} & = & {\psibar} ( i \delsla - M ) \psi
+ \frac{1}{2}\partial_{\mu}{\phi_a} \partial^{\mu}{\phi_a}
- \frac{1}{2} m_{\pi}^2 {\phi_a}{\phi_a}
\nonumber \\ 
& &
+ \frac{1}{2}\partial_{\mu}{\sigma} \partial^{\mu}{\sigma}
- {\widetilde U} [\sigma] 
- \frac{1}{4} {\omega_{\mu \nu}} {\omega^{\mu \nu}}
+ \frac{1}{2} {m_{\omega}^2} 
\omega_\mu \omega^\mu ,
\nonumber \\ 
& & 
+ i \frac{f_{\pi}}{m_{\pi}} 
{\psibar} \gamma_{5} \gamma^{\mu} {\tau}_{a} \psi \partial_{\mu} \phi_{a}
+ \gs \bar{\psi} \psi \sigma
- \gv \bar{\psi} \gamma_\mu \psi \omega^\mu
\label{Lag}
\end{eqnarray}
with
\begin{equation}
{\omega_{\mu \nu}} = 
\partial_{\mu}{\omega_\nu} - \partial_{\nu}{\omega_\mu} ,
\end{equation}
where $\psi$, $\phi$, $\sigma$ and $\omega$ are 
the nucleon, pion, sigma-meson and omega-meson fields, 
respectively.
In the above expression 
we use the pseudo-vector coupling
form as an interaction between Nucleon and pion.
The self-energy potential of the $\sigma$-field
${\widetilde U} [\sigma]$ is given as
\begin{equation}
\widetilde{U} [\sigma] 
= \frac{ \frac{1}{2} {m_{\sigma}^2} \sigma^2 
+ \frac{1}{3} B_\sigma \sigma^3
+ \frac{1}{4} C_\sigma \sigma^4 } 
{ 1 + \frac{1}{2} A_\sigma \sigma^2 } \  .
\eq{sigself}
\end{equation}
The symbols
$m_\pi$, \ms and \mv  are the masses of {$\pi$}-, {$\sigma$}- and
{$\omega$}-mesons, respectively.

Next we calculate the nucleon self-energies.
The nucleon self-energies are separated into the local part
and the momentum-dependent part as
\begin{equation}
U_{\alpha} (p) = U_{\alpha}^{L} + U_{\alpha}^{MD} (p).
~~~~~~~( \alpha = s, \mu )
\end{equation}
The {$\sigma$}- and {$\omega$}-meson exchange parts
produce
only very small 
momentum-dependence of nucleon self-energies \cite{RHF,soutome}
as their masses are large.
In fact the RH and RHF approximations
do not give any different results in nuclear matter properties 
after fitting parameters 
of {$\sigma$}- and {$\omega$}-exchanges \cite{RHF};
it is very easily confirmed in the limit of 
$m_{\alpha} \rightarrow \infty$ at the fixed 
${g_\alpha}/m_{\alpha}$.
On the other hand the one-pion exchange force is a long range 
one and makes large momentum-dependence while it does not
contribute to the local part in the spin-saturated system.
Thus we make the local part by RH
of the {$\sigma$}- and {$\omega$}-meson exchanges, and
the momentum-dependent part by RHF
of the pion exchange.
Please note that such a separated method can keep 
the consistency for the energy-momentum tensor \cite{KLW1}.

In this model the local part of the self-energies are given as
\begin{eqnarray}
U_{s}^{L} & = & \gs \sigm
\\
U_{\mu}^{L} & = & \delta_{0 \mu} \frac{\gv^2}{m_{\omega}^2} 
\rho_H
\end{eqnarray}
where $\sigm$ is the scalar mean-field obtained as
\begin{equation}
\frac{\partial}{\partial \sigm} {\tilde U} [\sigm]
= \gs \rho_s  
\end{equation}
In the above equations the scalar density $\rho_s$
and the vector Hartree density $\rho_H$ are given by
\begin{eqnarray}
\rho_s & = & 4	\int \frac{{\rm d}^3 {\vp}}{(2 \pi)^3}  
n({\vp}) 
\frac{M_{\alpha}^* (p)}{ \Pitild_0 (p) }   ,
\\
\rho_H & = & 4	\int \frac{{\rm d}^3 {\vp}}{(2 \pi)^3}  
n({\vp}) 
\frac{\Pi_0 (p)}{ \Pitild_0 (p)}   ,
\label{rhos}
\end{eqnarray}
where $n({\vp})$ is the momentum-distribution,
and  $\Pitild_{\mu} (p)$ is defined by
\begin{equation} 
\Pitild_{\mu} (p) = \frac{1}{2} \frac{\partial}{\partial p^{\mu}}
[ \Pi^2 (p) - {M^*}^2 (p) ]
\end{equation}
Please note that the Hartree density $\rho_H$ is not equivalent
to the baryon density $\rho_B$ when the self-energies depends
on the four momentum.

As a next step we define
the momentum-dependent parts of the self-energies as
the Fock parts with the one-pion exchange. 
When using the pseudo vector (PV) coupling 
the Fock parts do not become zero
at the infinite limit of the momentum $|\vp|$.
One usually erases these contributions by introducing 
the cut-off parameter.
In this work, instead of that, 
we minus these contributions from the momentum-dependent
parts (these contributions 
can be renormalized into the Hartree parts):
$U_\alpha \rightarrow U_\alpha - U_\alpha (p \rightarrow \infty)$.
Thus we obtain the momentum-dependent parts of the self-energies
as
\begin{eqnarray}
U_{s}^{MD} (p) & = &
\frac{3 \fpisq}{2}
\int \frac{d^3 \vk}{(2 \pi)^3} 
n({\vk}) 
\frac{M^* (k)}{ \Pitild_0 (k) } \Delta_{\pi} (p-k)   ,
\label{USPV}
\\
U_{\mu}^{MD} (p) & = & 
- \frac{3 \fpisq}{2}
\int \frac{d^3 \vk}{(2 \pi)^3} 
n({\vk}) 
\frac{\Pi_{\mu} (k)}{ \Pitild_0 (k) } \Delta_{\pi} (p-k)  ,
\label{UVPV}
\end{eqnarray}
where the ${\Delta_\pi}(q)$ is the pion propagator defined as
\begin{equation}
{\Delta_\pi}(q) = \frac{1}{q^2 - \mpisq} .
\end{equation}

In the above vector self-energies
we omit the tensor-coupling part  
involving $[ \Pi(k) \cdot  (p-k)] (p-k)_{\mu}$. 
This term is very small if the self-energy
is independent of momentum \cite{RHF}, and
their momentum-dependence is actually very small 
as shown later.

Using the above formulation we get the total energy density
of the spin-isospin saturated nuclear matter 
with the momentum-distribution $n_(\vp)$ as
\begin{eqnarray}
E_{T}/\Omega & = &  4 \int \frac{{\rm d}^3 {\vp}}{(2 \pi)^3} 
n({\vp}) {\varepsilon (\vp)}
+ {\widetilde U}[\sigma]
\nonumber \\ & &
+ ~ 2 \int \frac{{\rm d}^3 {\vp}}{(2 \pi)^3} n({\vp}) 
\frac{M^{*}(p) U_s^{MD}(p) - \Pi_{\mu}(p) U^{\mu}(p)}{ \Pitild_0 (p) } ,
\end{eqnarray}
with the system volume $\Omega$.

Let us show the calculated results in our model
using the momentum-dependent Dirac self-energies
with the one-pion-exchange.
In this calculation we fit the parameters (PF1) for 
the {$\sigma$}- and {$\omega$}- exchanges to reproduce 
the saturation properties that the binding energy
$BE = 16$MeV, the incompressibility $K = 200$MeV and
the effective mass $M^{*}/M = 0.7$ at the saturation
density $\rho_0 = 0.17$fm$^{-3}$.
For comparison we give the results with
the momentum-independent self-energies in
the parameter-set PM1 \cite{K-con} which gives
the same saturation properties. 
The detailed values of the parameter-sets PF1 and PM1 are
given in Table 1. 

In Fig. 1 we draw the momentum-dependence of
the scalar self-energy $U_s (p)$ and 
the time component of the vector self-energy $U_0 (p)$.
It can be seen that the variation of 
the momentum-dependent self-energies is only 2.5 \%
at most below Fermi level, which looks very small.

In Fig. 2 we show the density-dependence of the Dirac
self-energies $U_s$ and $U_0$ on the Fermi-surface (a)  
and the Landau mass (b) 
with the parameter-sets of PF1 and PM1.
Though two results of $U_s$ and $U_0$ almost agree 
together,
we can see rather large difference in the Landau mass: 
the value at $\rho_B = \rho_0$
is $M^{*}_L /M = 0.85$ in PF1, which is
consistent with the value expected by the analysis
of ISGQR as shown previously, while 
the momentum-independent calculation (PM1) gives
$M^{*}_L /M = 0.74$.
Hence it is shown that the very small momentum-dependence 
in the nucleon self-energies
enhances the Fermi velocity about 15 \%, and gives a 
significant difference in the Landau mass.
Furthermore we can also see an interesting behavior of 
$M^{*}_L$ in PF1, namely, its value agrees with the bare mass at 
$\rho_B \approx 0.5 \rho_0$ and becomes larger with the decrease
of the density.
Effects of small Dirac effective mass
are largely canceled at low density by the momentum-dependence created 
by the one-pion exchange.
This fact implies that the non-locality of the self-energies
affects nuclear surface properties such as
the isovector magnetic moment,
whose value is still larger than the Schmidt value \cite{magmom}.

Here we should give a further comment.
Bentz et al. have shown in Ref. \cite{Benz2} that the Landau mass 
is reduced by the one-pion exchange, which is opposite to ours.
In this calculation Benz et al. have used 
the pseudo-scalar (PS) coupling, and the sign of 
$U^{MD}_{\mu}$ was taken to be opposite to ours.

In order to understand this disagreement more we consider 
the Schr\"odinger equivalent 
potential only with the one-pion exchange.
In the non-relativistic limit the Schr\"odinger equivalent 
potential is given as
\begin{equation}
U_{SEP} = - U_s + U_0 + \frac{\vp^2}{2M} U_0.
\end{equation}
Both the PS and PV couplings make same results up to the first order 
of $M^{-2}$ and $m_\pi^{-2}$.
This agreement is caused by the fact
$U_s^{MD} - U_0^{MD} \approx 0$  in the PS coupling at its zeroth order,
while the proper PV coupling does not contribute to the zeroth order.  
However these two parts $U_s^{MD}$ and $U_0^{MD}$  are not
small in the PS coupling 
because the PS coupling constant $g_{\pi}$ is rather big.
In the relativistic mean-field framework, however,
the self-energies $U_s$ and $U_0$ have as large strengths as
the nucleon mass $M$, and then
the cancellation in the non-relativistic limit does not occur.
This fact is the origin of the difference between the two kinds of 
couplings.
The full HF calculation with 
the PS coupling makes too large contribution to 
the Dirac self-energies \cite{Tjon} while
Bentz el al. calculated the Fock term with the perturbative 
way.
Thus a calculation with the PV coupling must be more reliable 
than that with the PS coupling.

In this paper we have shown that the resonance energy of ISGQR
is given with the same formula as the non-relativistic macroscopic theory;
this value is determined by the Landau mass $M^{*}_L$.
The typical value of Dirac effective mass is empirically
known as $M^{*}/M = 0.55 - 0.7$, which corresponds to  
$M^{*}_{L}/M = 0.6 - 0.75$ in the RH approximation
with the momentum-independent Dirac self-energies. 
Such a small Landau mass overestimates the excited energy of ISGQR.
On the contrary, if we introduce the explicit momentum-dependences, 
the Fock part of the one-pion exchange enhances $M^{*}_L$. 
As an example, we have shown that  
Dirac effective mass $M^*/M = 0.7$ corresponds to the 
Landau mass $M^*_L/M = 0.85$, which is consistent with 
the experimental result of the ISGQR energy, which is in 
strong contrast to the value.
$M^*_L/M = 0.74$ in RH.

Of course we still have some ambiguities in this work. 
For example the bulk density of finite nuclei is smaller than
the saturation density, so that we may discuss the value
of the Landau mass at lower density.
In Ref.\cite{Vretenar}, furthermore, it has been reported that 
the time-dependent mean-field calculations have explained
the excited energies of ISGQR for $^{16}$O and $^{40}$Ca.
These results are inconsistent with the macroscopic theory. 
The treated nuclei may be too small
or their calculations involve other correlation beyond
the macroscopic theory.
Thus we should not make a quantitative conclusion 
on the ISGQR state before investigating it in finite nuclei.
Though we still have ambiguities, nevertheless, we can conclude 
that the momentum-dependence 
largely affects the Fermi-velocity, particularly 
in the low density region.

Here we should note that although the introduction of 
the momentum-dependence changes the Landau mass,
the depth of self-energies and hence 
the density dependence of the
total energy are affected very little.
Thus the approximation to neglect the non-locality of the
Dirac-field should be correct in discussions of many aspects of the
nuclear structure in the Dirac approach.
However, as for some physical quantities such as Fermi velocity, 
we will have to take account of the non-locality effects
in the Dirac approach.
This effect cannot be involved even if the density-dependent
parameters are introduced into the RH approximation 
\cite{hirata,Cheng}.

The authors would like to thank Prof. S. Nishizaki and 
Dr. Guangjun Mao for stimulating discussions on this work.


%

\newpage

\begin{center}
\begin{small}
\begin{tabular}{|c|ccccc|}
\hline \hline 
\stret{25pt}
 & \gs & \gv & $B_\sigma$ & $A_\sigma$ & $f_\pi$ \\
\hline 
\stret{20pt}
PF1 & 9.699 & 9.880 & 27.61 & 6.134 & 1.008 \\ 
\hline 
\stret{20pt}
PM1 & 9.408 & 9.993 & 23.52 & 5.651 & 0.0 \\
\hline
\end{tabular}

\vspace*{0.6cm}

\begin{minipage}{10.5cm}
\noindent
Parameter sets in this paper.
In all cases have used \mpi = 138 MeV, \ms = 550 MeV,
\mv = 783 MeV and $C_\sigma$ = 0.
\end{minipage}

\end{small}
\end{center}

\newpage

{\Large Figure Captions}

\bigskip

\begin{itemize}

\item[Fig. 1]
Momentum-Dependence of the scalar (a)
and vector (b) self-energies.
The solid  and dashed lines indicate
the results in PF1 and PM1, respectively.
The dotted line denotes the position of 
the Fermi momentum at $\rho_B = \rho_0$.

\item[Fig. 2]
Density-dependence of the Dirac
self-energies $U_s$ and $U_0$ on the Fermi-surface (a)  
and the Landau mass (b) the Landau mass (b).
The solid and dashed lines indicate the results for PF1 and PM1,
respectively, and the full square in (b) denotes the value
expected empirically from ISGQR.

\end{itemize}

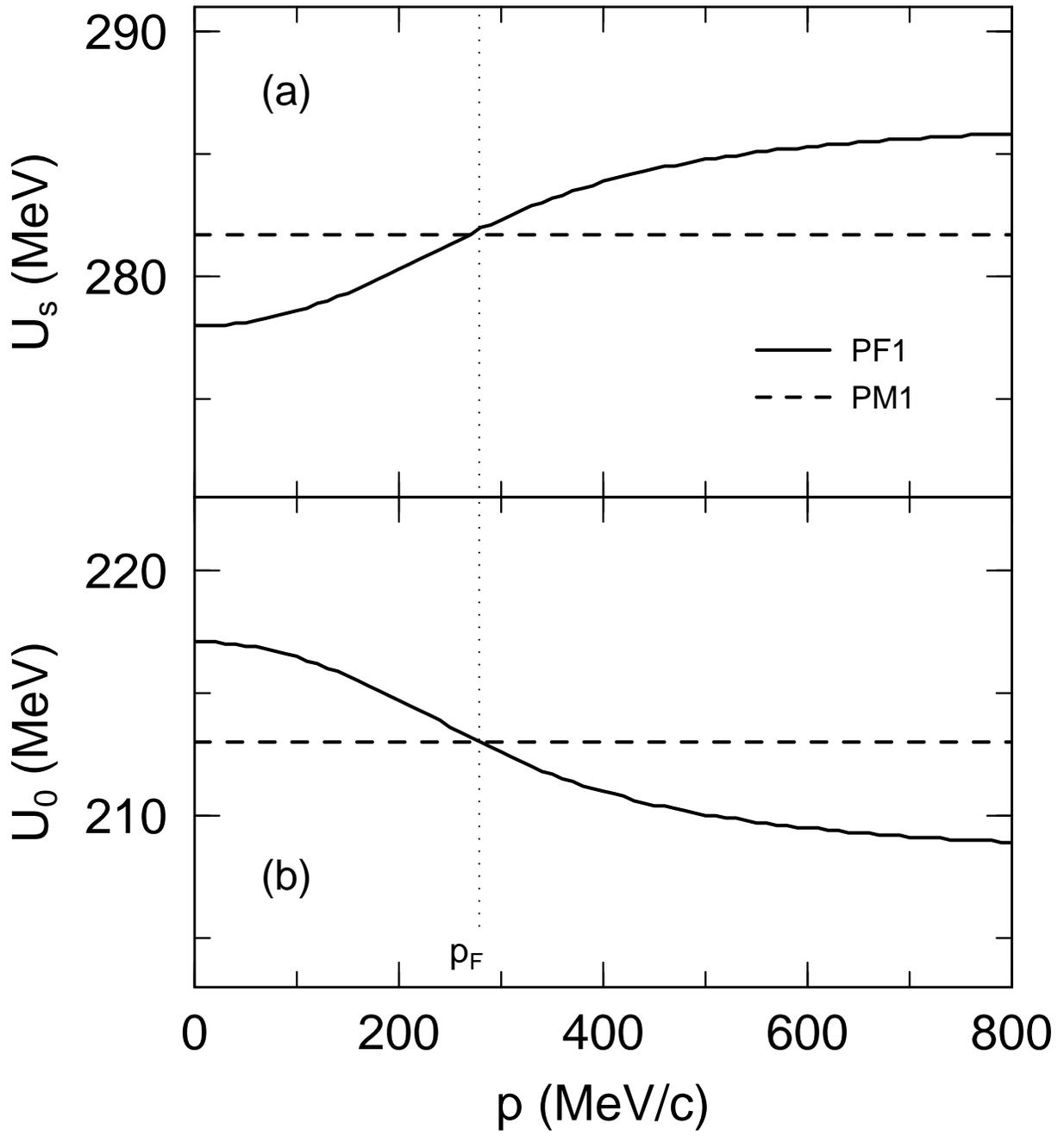
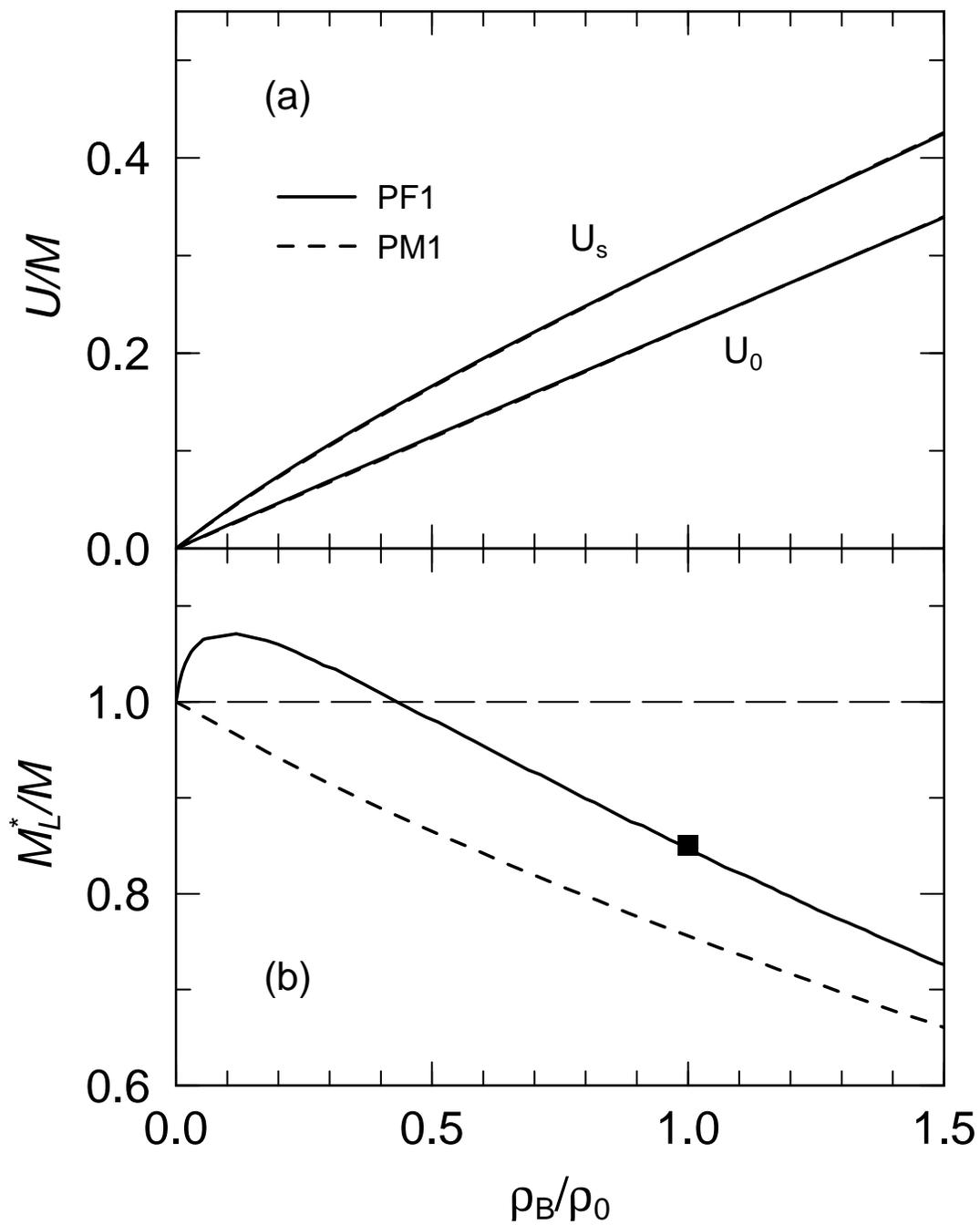
\end{document}